\documentclass[
reprint,
 amsmath,amssymb,
 aps,
prd,
]{revtex4-2}

\usepackage{dcolumn}
\usepackage{bm}
\usepackage[dvipsnames]{xcolor}
\usepackage{graphicx}
\usepackage[unicode]{hyperref}
\hypersetup{colorlinks=true, citecolor=MidnightBlue,
            linkcolor=MidnightBlue, urlcolor=MidnightBlue, linktocpage=true}
\usepackage{multirow}

\usepackage{array}
\usepackage{enumitem}
\usepackage{mathtools}

\begin{document}

\title{Can wormholes have vanishing Love numbers?}

\author{Shauvik Biswas}
\email{shauvikbiswas2014@gmail.com}
\affiliation{Department of Physics, Indian Institute of Technology, Guwahati 781039, India}

\begin{abstract}
    Wormholes are fascinating alternatives to black hole geometries. In this paper, we have studied a special case of wormhole solution in the context of $R=0$ spacetime. Our approximate analytical calculations show that under a strictly static axial gravitational perturbation of this spacetime, the magnetic-type tidal Love number (for $\ell=2$) vanishes if we keep the solution of the master equation up to linear order in the regularisation parameter of the geometry. Moreover, imposing physically viable boundary conditions on the master variable we have corroborated our result with the numerical solutions. We have further extended our analysis towards the case of Damour-Solodukhin wormhole, which also yields similar conclusion.
\end{abstract}
\maketitle
\section{Introduction}
Nowadays, gravitational wave astronomy astronomy of binary black hole merger\cite{LIGOScientific:2016aoc,LIGOScientific:2025rid} has opened the window to test general relativity in the previously unexplored strong and dynamic regime of gravity in an unprecedented manner\cite{Berti:2025hly,Berti:2024orb,Will:2014kxa,Berti:2015itd}. Such observations can give us scope not only to test general relativity but also to test the black hole nature of the compact objects\cite{Cardoso:2019rvt}. One way to quantify the nature of the compact object is to measure it’s tidal deformation during the inspiral stage, which can affect the orbital dynamics and the emitted gravitational waves (GWs)\cite{Binnington:2009bb}. The tidal deformability of a self-gravitating object is characterised by tidal Love numbers (TLNs)\cite{wilson1914aeh, Murray_Dermott_2000}. It relates the induced multipole moments to the applied external tidal field \cite{PhysRevD.98.124023,Cardoso:2017cfl}. Therefore, the tidal deformability of the body encodes information regarding the body’s internal structure\cite{Cardoso:2017cfl}. In more mathematical terms, if a spherically symmetric star of mass $M$ is placed in an external quadrupolar static tidal field $\mathrm{E}_{ij}$ and in response to it, the star develops a mass quadrupole moment $Q_{ij}$, then one can show that to linear order in $\mathrm{E}_{ij}$\cite{Hinderer:2007mb},
\begin{align}
    Q_{ij}=-\lambda~\mathrm{E}_{ij}~.
\end{align}
Here, the proportionality constant $\lambda$ represents the tidal Love number. In some recent works\cite{Chia:2020yla,Chakrabarti:2013xza,Goldberger:2004jt}, it has been shown that in a dynamic situation, the tidal response of a body in an external tidal field can be decomposed into two parts\cite{Poisson_Will_2014}. The first part of which represents the deformation of the body and is conservative. The other part of the tidal response represents the absorption of the GWs by the body and is dissipative. The tidal Love number characterises the conservative part. Similarly, there are also tidal dissipation numbers.~TLNs of compact objects affect the phase of the gravitational waveform of the binary at a higher post-Newtonian order\cite{PhysRevD.77.021502,Bini:2012gu}(5PN), and the tidal dissipation number affects the waveform at 2.5PN order\cite{PhysRevD.87.044022,Porto:2007qi}. However, for non-rotating bodies, the tidal dissipation number first enters in the phase of the GW waveform at 4PN order.\\
\indent Surprisingly, in the four dimensions the TLNs of the asymptotically flat black holes (under bosonic perturbation, see \cite{Chakraborty:2025zyb} for the fermionic case) within the context of general relativity (GR) are exactly zero\cite{Binnington:2009bb} in the static scenario\cite{Chakraborty:2023zed}. Such a precise cancellation of the TLNs within the context of GR seems to pose a ``naturalness" problem\cite{Porto:2016pyg,Porto:2016zng} similar to that which appears in the standard model of particle physics\cite{Veltman:1980mj}. One way to cure the problem is by arguing that GR is not a correct theory of gravity near the black hole horizon, and one should look for other alternative theories of gravity. On the other hand, since black hole evaporation (due to Hawking radiation) violates quantum unitarily\cite{PhysRevD.14.2460}, one argues that gravitational collapse should not lead to black holes but to horizonless compact objects\cite{Mathur:2008kg} with a photon sphere so that they can mimic observations with black holes. In literature, these compact objects are known as exotic compact objects (ECOs)\cite{Cardoso:2019rvt}. Interestingly, they have a non-vanishing Love number, which can be used to distinguish them from general relativistic black holes. Several studies have been performed to compute the tidal Love number of neutron stars\cite{De:2018uhw,LIGOScientific:2018cki}, ultra compact objects\cite{Yagi:2016ejg,Yazadjiev:2018xxk,Chakravarti:2019aup}, black holes in alternative\cite{DeLuca:2022tkm,Nair:2022xfm,Tan:2020hog} and higher dimensional theories of gravity\cite{Singha:2025xah,Charalambous:2023jgq,Rodriguez:2023xjd}, black holes in asymptotically non-flat spacetime\cite{Franzin:2024cah,Emparan:2017qxd} and BTZ black holes\cite{Bhatt:2024mvr}. Similarly, the tidal responses of the regular black holes are also studied in\cite{Coviello:2025pla}.\\
\indent In the case of non-spinning compact objects, depending on the parity of the perturbation, tidal Love number can be of two types\cite{Binnington:2009bb}: a) electric-type and b) magnetic-type. The induced mass multipole moments are associated with the electric-type tidal Love number. On the other hand, the current multipole moments are associated with magnetic type tidal Love number. These multiple moments are generated due to an external magnetic type (odd parity) tidal field. Since in the Newtonian gravity, there is no odd parity sector, there is no Newtonian analogue of these Love numbers. This was first time introduced by Damour et al.\cite{PhysRevD.45.1017} and Favata\cite{Favata:2005da} in the post-Newtonian theory. Moreover, they have smaller effects on the GW waveform from the binary as compared to the electric ones.\\
\indent In this letter, we will be interested in considering wormhole spacetime\cite{PhysRev.128.919,PhysRev.48.73} as an ECO\cite{Biswas:2022wah} and computing the static magnetic Love number associated with them. In particular, we will consider $R=0$ spacetime and a special class of wormhole solutions within the context of general relativity\cite{Dadhich:2001fu}. Historically, it has been shown that in the context of the braneworld scenario, such a wormhole solution \textit{does not} require exotic matter\cite{Kar:2015lma} to sustain it. Moreover, they are shown to be stable under linear perturbations and therefore to be astrophysically viable\cite{Biswas:2022wah,Biswas:2023ofz}.~We will further extend our analysis towards the case of Damour-Solodukhin wormhole.\\

\indent The remainder of the letter is organised as follows: In Sec.\ref{sec-2}, we have given the brief description of the wormhole solution in the context of $R=0$ spacetime. In Sec.\ref{sec-3}, we have discussed the axial gravitational perturbations of an anisotropic fluid star and the associated strictly static limit. Sec.\ref{sec-4} contains the main conclusion of the paper, where we have solved the master equation perturbatively. Sec.\ref{sec-5} contains numerical investigations and comparison with the analytical case. In Sec.\ref{sec-6} we have extended our analytical computations towards the Damour-Solodukhin wormhole. We end our discussion with a concise conclusion in Sec.\ref{sec-7}.\\
\indent Throughout the paper, we will use geometrized units, that is we will set $c=G=1$.

\section{Brief description of the $R=0$ Wormhole Spacetime}\label{sec-2}
In order to model a static spherically symmetric wormhole spacetime, one used to take the following metric ansatz,
\begin{align}\label{wormhole metric ansatz}
    ds^{2}=-e^{2\Phi(r)}dt^{2}+\frac{dr^{2}}{1-\frac{b(r)}{r}}+r^{2}\left(d\theta^{2}+\sin^{2}\theta d\phi^{2}\right)~.
\end{align}
Here, $\Phi(r)$ is the redshift function and $b(r)$ is the shape function. The geometrical requirement of traversability and that the spacetime connects two asymptotically flat spacetimes by a bridge like structure (throat) imposes the following mathematical conditions on the metric functions\cite{visser1995lorentzian,Morris:1988cz}:
\begin{itemize}
    \item $e^{2\Phi(r)}$ must not have zeros. Often this condition is referred to as the ``no-horizon" condition.
    \item existence of wormhole throat ($r=r_{0})$ implies that $b(r_{0})=r_{0}$ such that $\frac{b(r)}{r}\leq 1~\forall~r\geq r_{0}$.
    \item asymptotic flatness implies that $\Phi(r\rightarrow\infty)=0$ and $\frac{b(r)}{r}\rightarrow 0$ as $r\rightarrow 0$.
\end{itemize}
If one further assumes that the spacetime is sourced by an anisotropic fluid, whose energy momentum tensor has the following form $T^{\mu}{}_{\nu}=\rm\text{diag}(-\rho,p_{r},p_{t},p_{t})$, then together with the metric, we have five unknowns. However, considering Einstein's gravity, we have three equations corresponding to the equations associated with $G^{t}{}_{t}$, $G^{r}{}_{r}$ components and $\nabla_{\mu}T^{\mu}{}_{r}=0$. So, one needs two more conditions to construct the wormhole spacetime. Dadhich et al.\cite{Dadhich:2001fu} considered these two conditions to be $\rho=0$ and $R=0$. Using Einstein's equation, it can be shown that the first condition implies $b=\text{constant}\equiv 2M$. Exploiting this condition, one can solve the $R=0$ equation to get,
\begin{align}\label{Dadhich}
  g_{tt}=-\left(\kappa+\alpha\sqrt{1-\frac{2M}{r}}\right)^{2}~,  
\end{align}
where $\kappa$ and $\alpha$ are two integration constants. Further, one can rescale the time coordinate $t\rightarrow\frac{t}{(\kappa+\alpha)}$ to re-cast $g_{tt}$ as,
\begin{align}\label{g-tt}
  g_{tt}=-\frac{1}{(p+1)^{2}}\left(p+\sqrt{1-\frac{2M}{r}}\right)^{2}~,\quad p=\frac{\kappa}{\alpha}.   
\end{align}
From the perspective of astrophysical interest, we require this spacetime to be a good mimicker of Schwarzschild spacetime, which demands $p\gtrsim 0$. Before we proceed to the next section, let us put the expressions of the components of the energy momentum tensor,
\begin{align}\label{TEM}
 \rho=0,~&p_{r}=-\frac{2 p M}{8\pi r^{3}\left(p+\sqrt{1-\frac{2M}{r}}\right)},\\
 &~p_{t}=\frac{p M}{8\pi r^{3} \left(p+\sqrt{1-\frac{2M}{r}}\right)}~.   
\end{align}
From the expression of the energy momentum tensor components, it is clear that the wormhole spacetime violates the weak and null energy conditions which can be connected with the geodesic defocusing near the throat\cite{Shaikh:2018yku,PhysRev.98.1123}. Interestingly, even though in GR this spacetime violates the energy conditions, it was shown that in the braneworld scenario, due to the existence of the radion field on the brane, one can cure the violation of the energy conditions.  
\section{Axial Gravitational Perturbation of Anisotropic Fluid Star and the static limit}\label{sec-3}
On top of the background spacetime $g^{(0)}_{\mu\nu}$, we put the perturbations $h_{\mu\nu}$, such that 
\begin{align}
g_{\mu\nu}=g^{(0)}_{\mu\nu}+h_{\mu\nu}~\text{with}~|h_{\mu\nu}|\ll |g^{(0)}_{\mu\nu}|~.
\end{align}
Since the background spacetime is static and spherically symmetric, the perturbations $h_{\mu\nu}$ can be decomposed into the radial, temporal, and angular parts by using tensor spherical harmonics. 
Based on the parity transformation properties of the tensor spherical harmonics, we can decompose $h_{\mu\nu}$ into the axial and polar sectors. Among the ten independent components of the symmetric tensor $h_{\mu\nu}$, the axial sector contains three, and the polar sector contains seven degrees of freedom. Now, due to the diffeomorphism invariance of the theory, we can impose gauge conditions on $h_{\mu\nu}$ such that, $h_{\theta\phi}=0$\cite{PhysRev.108.1063}. In that case, the axial perturbations are described by the following components\cite{PhysRev.108.1063},
\begin{align}\label{pert_grav_ax}
h_{\mu\nu}=e^{-i\omega t}
\left(
\begin{array}{cccc}
0 & 0 & -\frac{h_{0}(r)}{\sin \theta}\partial_{\phi}Y_{\ell m} & h_{0}(r)\sin \theta \partial_{\theta}Y_{\ell m}\\
0 & 0 & -\frac{h_{1}(r)}{\sin \theta}\partial_{\phi}Y_{\ell m} & h_{1}(r)\sin \theta \partial_{\theta}Y_{\ell m} \\
\textrm{sym} & \textrm{sym} & 0& 0 \\
\textrm{sym} & \textrm{sym} & 0 & 0\\
\end{array}
\right).
\end{align}
Here $Y_{\ell m}(\theta,\phi)$ is the scalar spherical harmonics. However, because of spherical symmetry of the background spacetime, the obtained master equation will be independent of $m$ and then it is helpful to set $m=0$ from the beginning. Therefore we will have only the following non-vanishing components:
\begin{align}\label{components-h-mu-nu}
  h_{t\phi}=h_{0}(r)\sin\theta \frac{dP_{\ell}}{d\theta}e^{-i\omega t}~,\\
  h_{r\phi}=h_{1}(r)\sin\theta \frac{dP_{\ell}}{d\theta}e^{-i\omega t}~.
\end{align}
Up to linear order in $h_{\mu\nu}$ perturbed components of the Einstein tensor in the axial sector are
\cite{Kojima:1992ie,Chakraborty:2024gcr}\footnote{Here we have used $P_{\ell}$ to denote $P_{\ell}(cos\theta)$.},
\begin{align}\label{G-r-Phi}
    &\delta G_{r\phi}=\delta R_{r\phi}-R\delta g_{r\phi}\nonumber\\
   &=\frac{e^{-i\omega t}}{2}\sin\theta\frac{dP_{\ell}}{d\theta}\bigg[-\frac{\omega^{2}}{f}h_{1}+\frac{i\omega}{f}r^{2}\frac{d}{dr}\left(\frac{h_{0}}{r^{2}}\right)\nonumber\\
   &-h_{1}\{\frac{g^{\prime}}{r}+\frac{g}{r^{2}f}(2f+rf^{\prime})\}+h_{1}\frac{\ell(\ell+1)}{r^{2}}\bigg]\nonumber\\
   &-h_{1}e^{-i\omega t}\sin\theta \frac{d P_{\ell}}{d\theta}\bigg[\frac{2}{r^{2}}\bigg(1-\frac{1}{g}\bigg)+\frac{2}{rg}\bigg(\frac{g^{\prime}}{g}-\frac{f^{\prime}}{f}\bigg)\nonumber\\+&\frac{f^{\prime}}{2 g f}\bigg(\frac{f^{\prime}}{f}+\frac{g^{\prime}}{g}\bigg)-\frac{f^{\prime\prime}}{f g}\bigg]~;
\end{align}
\begin{align}\label{G-t-phi}
    &\delta G_{t\phi}=\frac{e^{-i\omega t}}{4 r^{2} f^{2}}\sin\theta\frac{dP_{\ell}}{d\theta}\bigg[2(\ell-1)(\ell+2)f^{2}h_{0}+\nonumber\\& g(4f^{2}h_{0}-4i r\omega f^{2}h_{1}+ir^{2}\omega ff^{\prime}h_{1}-r^{2}h_{0}f^{\prime}{}^{2}+r^{2}ff^{\prime}h_{0}^{\prime}\nonumber\\&-2i\omega r^{2}f^{2}h_{1}^{\prime}
+2r^{2}ff^{\prime\prime}h_{0}-2r^{2}f^{2}h_{0}^{\prime\prime})\nonumber\\&+g^{\prime}(4rf^{2}h_{0}-ir^{2}\omega f^{2}h_{1}
+r^{2}ff^{\prime}h_{0}-r^{2}f^{2}h^{\prime}_{0})\bigg]~~\text{and}
\end{align}
\begin{align}
 &\delta G_{\theta\phi}=\frac{1}{2}e^{-i\omega t}\sqrt{\frac{g}{f}}\bigg[-i\omega\frac{h_{0}}{\sqrt{fg}}-\frac{d}{dr}\bigg(\sqrt{fg}h_{1}\bigg)\bigg]\nonumber\\
 &\bigg[\ell(\ell+1)\sin\theta P_{\ell}+ 2\cos\theta \frac{dP_{\ell}}{d\theta}\bigg]~.  
\end{align}
On the other hand, the non-vanishing perturbed components of the energy momentum tensors in the axial sector are,
\begin{align}
    \delta T_{r\theta}=p_{t}h_{1}e^{-i\omega t}\sin\theta\frac{dP_{\ell}}{d\theta}~,
\end{align}
 \begin{align}
  \delta T_{t\phi}=-\bigg(h_{0}\rho+\frac{f(r)a(r,\omega)}{4\pi}\bigg)e^{-i\omega t}\sin\theta \partial_{\theta}P_{\ell}
 \end{align}
where $a(r,\omega)$ is related to the perturbed component of the fluid four-velocity in the following way\footnote{In general, there can be terms proportional to $\partial_{\phi}Y_{\ell m}$, but due to spherical symmetry we take $m=0$ for simplicity.},
\begin{align}
    \delta u^{\mu}= \frac{\sqrt{f} a(r,\omega)}{4\pi r^{2}(\rho+p_{t})}\frac{\partial_{\theta}P_{\ell}}{\sin\theta}\delta^{\mu}_{\phi}~.
\end{align}
In this paper, we are interested in the strictly static case of the perturbed Einstein's equation, which is described by the condition\cite{Pani:2018inf}:
\begin{align}\label{staticity}
    \omega=0=h_{1}=a(r,0)~.
\end{align}
That is, the static case of the axial perturbation is described by $h_{t\phi}$ only (see Eq.[\ref{components-h-mu-nu}]). Now using the perturbed Einstein's equation $\delta G_{\mu\nu}=8\pi\delta T_{\mu\nu}$ and using the condition Eq.[\ref{staticity}] we get, 
\begin{align}\label{master-Love}
    &g h_{0}^{\prime\prime}-4\pi r (\rho+p_{r})h_{0}^{\prime}-\nonumber\\
    &h_{0}\bigg[\frac{(\ell+2)(\ell-1)}{r^{2}}+8\pi(\rho-p_{r}+2p_{t})+\frac{2 g}{r^{2}}\bigg]=0~.
\end{align}
Instead of taking $u^{\mu}$ as a fundamental quantity for the fluid perturbation, if we had taken the fluid four-velocity to be hypersurface orthogonal, that is, $u_{\mu}$ as a fundamental quantity, then one can show that in the static case, we have a different master equation\cite{Chakraborty:2024gcr},
\begin{align}\label{master-Love-2}
   &g h_{0}^{\prime\prime}-4\pi r (\rho+p_{r})h_{0}^{\prime}-\nonumber\\
    &h_{0}\bigg[\frac{(\ell+2)(\ell-1)}{r^{2}}-8\pi(\rho+p_{r})+\frac{2 g}{r^{2}}\bigg]=0~.  
\end{align}
Note that even for the isotropic fluid case these two master equations are different and only in the vacuuum case they will be the same.
In the next section, we will solve the above two master equations perturbatively and show that if such a solution has to be regular at the wormhole throat, then the wormhole spacetime will have vanishingly small static magnetic tidal Love number.
\section{Vanishingly small static tidal Love number}\label{sec-4}
As we have argued previously, we need $p\gtrsim 0$, so that the wormhole spacetime behaves as a black hole mimicker. This requirement gives us the freedom to solve Eq.[\ref{master-Love}] in a perturbative manner, that is, we demand that the master variable $h_{0}(r)$ be of the form,
\begin{align}\label{h0-expansion}
    h_{0}(r)=\sum_{n=0}^{\infty}p^{n}h_{0}^{(n)}(r)~.
\end{align}
In what follows, we will demand that each term of the above expansion is regular in the wormhole throat, so that the entire solution $h_{0}(r)$ is regular there.
For simplicity, in this letter we will consider terms up to linear order. Plugging Eq.[\ref{h0-expansion}] in Eq.[\ref{master-Love}] and keeping terms up to linear order in $p$, we obatin
\begin{align}\label{linearzed-master}
    \mathcal{O}[h_{0}^{(0)}]+p \mathcal{O}[h_{0}^{(1)}]=p S[r]~,
\end{align}
where the differential operator $\mathcal{O}$ has the form,
\begin{align}\label{Operator-O}
    \mathcal{O}\equiv\left(1-\frac{2M}{r}\right)\frac{d^{2}}{dr^{2}}+\left(\frac{4M}{r^{3}}-\frac{\ell(\ell+1)}{r^{2}}\right)
\end{align}
and the inhomogeneous term on the right hand side of Eq.[\ref{linearzed-master}] is given by,
\begin{align}\label{linear-source}
    S[r]\equiv \frac{M}{r^{3}\sqrt{1-\frac{2M}{r}}}\left[4 h_{0}^{(0)}-r \frac{dh_{0}^{(0)}}{dr}\right]~.  
\end{align}
Then we get two equations at our disposal,
\begin{align}
    \mathcal{O}[h_{0}^{(0)}]=0~,\label{unper}\\
    \mathcal{O}[h_{0}^{(1)}]=S[r]\label{order-p}~.
\end{align}
In this paper, we will be interested in solving the above equations for the $\ell=2$ case. This has importance in the context of GW signals from compact binaries\cite{Favata:2005da,Flanagan:2007ix,Tichy:1999pv}. In particular, it is shown in \cite{Yagi:2013sva} the $\ell=2$ magnetic TLN enters in the GW phase at $6\rm PN$ order and the higher multipoles will affect the waveform at higher $\rm PN$ terms.

Imposing a regularity condition on the wormhole throat, one can show that the solution of Eq.[\ref{unper}] becomes\cite{Chakraborty:2026qru,Rodriguez:2026iot},
\begin{align}\label{h00}
h_{0}^{(0)}(r)= Ar^{3}\bigg(1-\frac{2M}{r}\bigg)~.    
\end{align}
where $A$ is an arbitrary constant.
Now using this solution in Eq.[\ref{linear-source}] and then inserting it in Eq.[\ref{order-p}] one will obtain a second order differential equation involving $h_{0}^{(1)}(r)$. However, one cannot solve it exactly in the presence of the inhomogeneous term $S[r]$. Now, from the perspective of the observations, $r\gg 2M$ is a regime of interest, so we approximate $S[r]$ as follows,
\begin{align}\label{approx-S}
     S[r\gg 2M]\sim AM-\frac{3 A M^{2}}{r}-\frac{5 M^{3} A}{2 r^{2}}-\frac{7 M^{4} A}{2r^{3}}+\mathcal{O}(\frac{1}{r^{4}})~.
\end{align}
Using this approximation in Eq.[\ref{order-p}] we get the following solution of $h_{0}^{(1)}(r)$,
\begin{align}\label{approx-h01}
   & h_{0}^{(1)}(r)=
\frac{1}{96 M^5 r}\bigg[-3 r^{3}\gamma (2 M-r) \log \left(1-\frac{2M}{r}\right)\nonumber\\
&-2 M \bigg(42 A M^7 r+93 A M^6 r^2+(29 A + 96 d_{1}) M^5 r^3\nonumber\\
&-(67 A +48 d_{1}) M^4 r^4-8 d_{2} M^3-8 d_{2} M^2 r\nonumber\\&-12 d_{2} M r^2+12 d_{2} r^3\bigg)\bigg]~.
\end{align}

Here $\gamma\equiv\left(39 A M^5-4 d_{2}\right)$ and $d_{1}$, $d_{2}$ are arbitrary constants associated with the differential equation of $h_{0}^{(1)}$.
Similarly, one can have an approximate analytical solution (see Appendix.\ref{app-A}) in the near throat region $r\gtrsim 2M$, and both solutions should be matched in the intermediate region given by $2M<r<\infty$. However, the regularity of such solutions at the throat of the wormhole dictates that the coefficient of the logarithmic term should vanish. It implies from Eq.[\ref{approx-h01}] that,
\begin{align}\label{Un-natural}
      39 A M^{5} = 4 d_{2}~.
\end{align}
Then the total solution becomes the following:
\begin{align}\label{h0-up-analytical}
    h_{0}&= p \bigg(\frac{\left(-84 A M^3-186 A M^2 r-58 A M r^2+134 A r^3\right)}{96}\nonumber\\
   & +\frac{13 A \left(4 M^4+4 M^3 r+6 M^2 r^2-6 M r^3\right)}{32 r}\bigg)\nonumber\\
   +&(A+p d_{1}) r^3 \left(1-\frac{2 M}{r}\right)~.
\end{align}
Now expanding, these total solution $h_{0}=h_{0}^{(0)}+p h_{0}^{(1)}$ for large values of $r$, one gets the following expression
\begin{align}\label{h0-large}
    h_{0}\sim&\frac{13 A M^4 p}{8 r}+\frac{3}{4} A M^3 p+\frac{1}{2} A M^2 p r\nonumber\\
    &+r^2 \left(-\frac{1}{24} (73 A +48 d_{1}) M p-2 A M\right)\nonumber\\
    &+r^3 \left(\frac{1}{48} (67 A +48 d_{1}) p+ A\right)~.
\end{align}
Following \cite{RevModPhys.52.299}, we recall that for asymptotically flat spacetime the metric component $g_{t\phi}$ admits the following large $r$ behavior:
\begin{align}\label{Thorne-Multipole}
    {g_{t\phi}}&=\frac{2J}{r}\sin^{2}\theta+\sin \theta \sum_{\ell\geq 2}\bigg[\frac{2}{r^{\ell}}\left\{\sqrt{\frac{4\pi}{2\ell+1}}\frac{\mathbb{S}_{\ell}}{\ell}\partial_{\theta}Y_{\ell 0}+(\ell'<\ell)\right\}\nonumber\\
    &+\frac{2r^{\ell+1}}{3\ell(\ell-1)}\left\{\mathbb{B}_{\ell}\partial_{\theta}Y_{\ell 0}+(\ell'<\ell)\right\}\bigg]~.
\end{align}
Here $\mathbb{S}_{\ell}$ is the induced spin multipole moment due to magnetic part $\mathbb{B}_{\ell}$ of the applied tidal field. Also $J$ denotes the angular momentum of the spacetime and $J=0$ for static spacetime. And $(\ell'<\ell)$ contains  terms of the form $\frac{1}{r^{\ell'}}$ with $\ell'<\ell$. This shows that at $r\rightarrow\infty$, $r^{\ell+1}$ is the dominant growing part and $r^{-\ell}$ will be dominant decaying part.
Then considering the above expression (see also \cite{Damour:2009vw,Cardoso:2017cfl,PhysRev.108.1063}), the general behavior of the $\ell=2$ mode for large $r$ can be divided into two parts: the growing part (which actually represents the applied external tidal field) goes as $r^{3}$, and the decaying part, which represents the body's response (here the wormhole throat) goes as $r^{-2}$.
In particular for $\ell=2$ the dimensionless magnetic TLN is defined as\cite{Chakraborty:2026qru,Cardoso:2017cfl},
\begin{align}\label{Magnetic-TLN2}
    k_{2}^{\rm B}\equiv -\frac{1}{M^{5}}\sqrt{\frac{4\pi}{5}}\frac{\mathbb{S}_{2}}{\mathbb{B}_{2}}.
\end{align}\\
\indent Comparing with Eq.[\ref{h0-large}] we see that if we set $A=0$ then the expression of $h_{0}(r\rightarrow \infty)$ matches with Eq.[\ref{Thorne-Multipole}] for $\ell=2$. One can easily check from Eq.[\ref{h0-up-analytical}] that it means,
\begin{align}\label{Bh-like-bs}
   h_{0}(r=2M)=0~. 
\end{align}
This is exactly the boundary condition obeyed in the black hole case (consider Eq.[\ref{h00}] for the case of black holes). Therefore along with this boundary condition and the regularity condition Eq.[\ref{Un-natural}], we see that the coefficient of $\frac{1}{r^{2}}$ term vanishes and we must conclude that the wormhole spacetime has a vanishingly small axial tidal Love number in the static (strictly) scenario. Thanks to the regularity condition at the throat. In the next section we will numerically analyze the consequences of any deviation from such boundary condition. For a purpose of sanity check of this conclusion in Appendix.\ref{app-B} we have studied the scalar perturbation.\\
Similarly, if we had considered that the fluid velocity is hypersurface orthogonal, then up to linear order in $p$ we get the following equation,
\begin{align}\label{linear-master-2}
    \mathcal{O}[h_{0}^{(0)}]+p \mathcal{O}[h_{0}^{(1)}]=p \tilde{S}[r]~.  
\end{align}
In this case, the source term in the right hand side of Eq.[\ref{linear-master-2}] becomes,
\begin{align}\label{linear-source-2}
    \tilde{S}[r]\equiv \frac{M}{r^{3}\sqrt{1-\frac{2M}{r}}}\left[2 h_{0}^{(0)}-r \frac{dh_{0}^{(0)}}{dr}\right]~.  
\end{align}
The solution for $h_{0}^{(0)}$ is already quoted in Eq.[\ref{h00}]. On the other hand, the solution for $h_{0}^{(1)}$ differs from the previous case and is given by,
\begin{align}\label{}
    &h_{0}^{(1)}=
    r^{3}\left(1-\frac{2M}{r}\right)\left(c_{1}+\frac{A}{12}\right)\log\left(1-\frac{2M}{r}\right)\nonumber\\
   & +\frac{c_{2}}{24 M^{5}r}(4 M^{4}+4 M^{3}r+6M^{2}r^{2}-6M r^{3})\nonumber\\
   & +\frac{c_{2}r^{2}}{4M^{4}}\left(1-\frac{r}{2M}\right)\log\left(1-\frac{2M}{r}\right)+\frac{A}{72}r^{2}(17 r-4M)~.
\end{align}
Here $c_{1}$ and $c_{2}$ are the two arbitrary constants associated with the differential equation ${O}[h_{0}^{(1)}]=\tilde{S}[r]$.
With the previous reasoning, if $h_{0}^{(1)}$ has to be regular at $r=2M$, then we must set $c_{2}=\frac{2}{3}AM^{5}$. Then the total solution $h_{0}=h_{0}^{(0)}+ph_{0}^{(1)}$ takes the following 
form,
\begin{align}\label{h0-down-analytical}
   h_{0}=&(A+p c_{1}) r^3 \left(1-\frac{2 M}{r}\right)+\frac{ p A r^2 (17 r-4 M)}{72}\nonumber\\
   &+\frac{p A \left(4 M^4+4 M^3 r+6 M^2 r^2-6 M r^3\right)}{36 r}~,
\end{align}
which has the following asymptotic form,
\begin{align}\label{h0-large-2}
  & h_{0}\sim r^{3} \left(p (\frac{17 A}{72}+c_{1})+A\right)\nonumber\\
  &-r^{2} \left(p(\frac{2 A M}{9}+2c_{1} M)+2 A M\right)\nonumber\\
   &+\frac{1}{6} A M^2 p r+\frac{1}{9} A M^3 p+\frac{A M^4 p}{9 r}~.
   \end{align}
Since this expression does not contain any $\frac{1}{r^{2}}$ term, so in this case also (along with the boundary condition Eq.[\ref{Bh-like-bs}], that is imposing A=0) the tidal Love number vanishes.
\section{Numerical Investigation}\label{sec-5}
\begin{figure}[htbp!]
\includegraphics[width=0.9\linewidth]{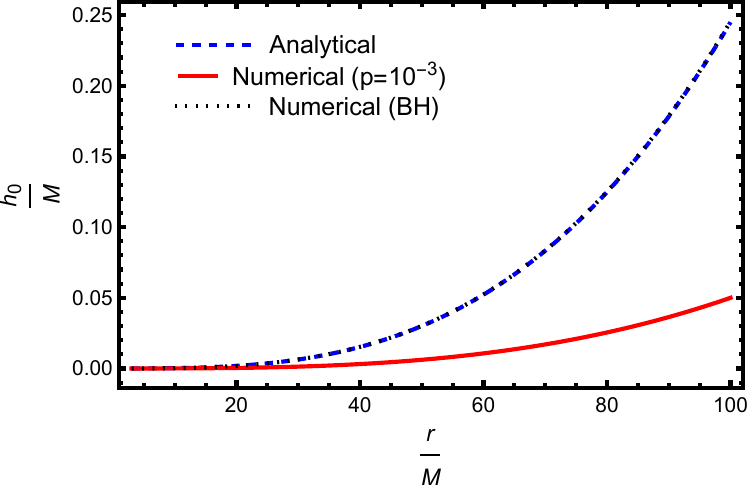}\\
\includegraphics[width=0.9\linewidth]{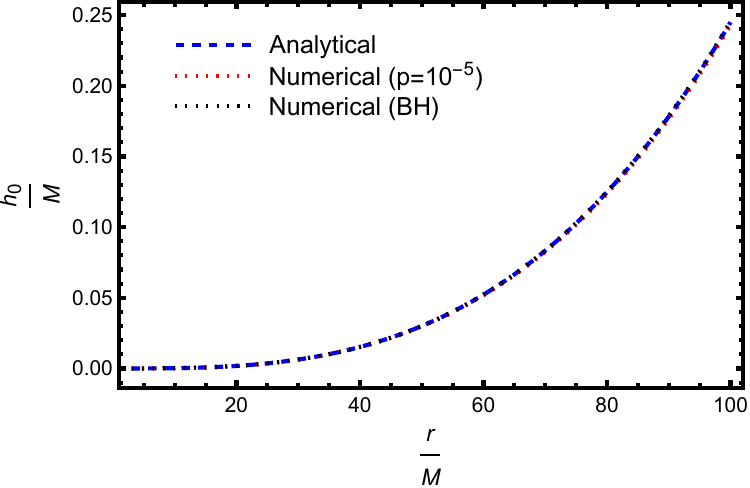}\\
\includegraphics[width=0.9\linewidth]{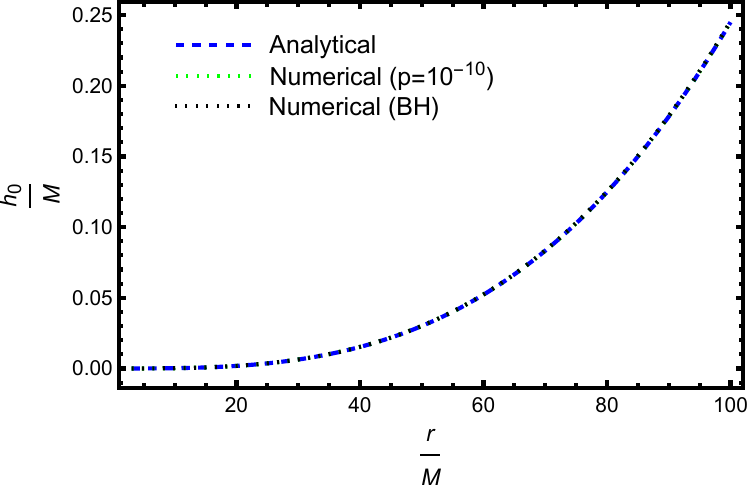}
\caption{In this figure we have checked if our analytical computation matches with full numerical solution of Eq.[\ref{master-Love-2}] for the case of $R=0$ spacetime. Here we have plotted $h_{0}(r)$ as a function of $r$ for both the cases of analytical and numerical computations and compared the result with the numerical solution of the Schwarzschild black hole cases. In the \textit{top panel} we have shown the comparison of analytical and numerical computations of $h_{0}$ for $p=10^{-3}$. It is clear that for large values of $r$, there is a huge difference. In the next two figures we have made the comparison for $p=10^{-5}$ and $p=10^{-10}$ respectively. It is clear that the exact numerical result matches well with the analytical result Eq.[\ref{h0-down-analytical}].}\label{Numerical-investigation-down}
\end{figure}
\begin{figure}[htbp!]
\includegraphics[width=0.9\linewidth]{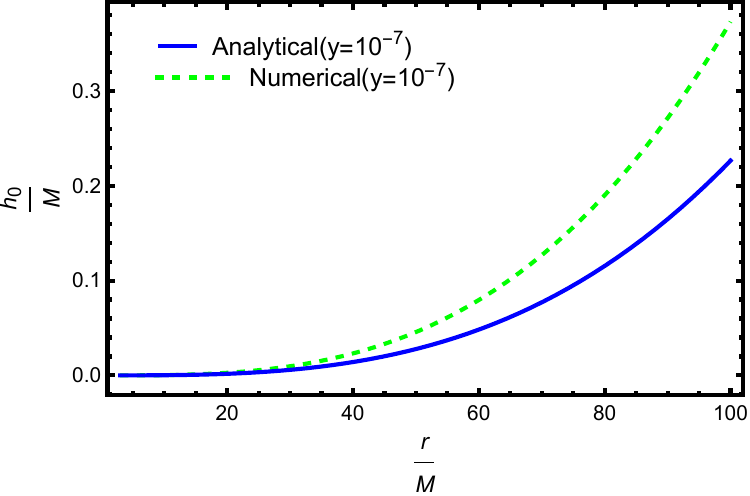}\\
\includegraphics[width=0.9\linewidth]{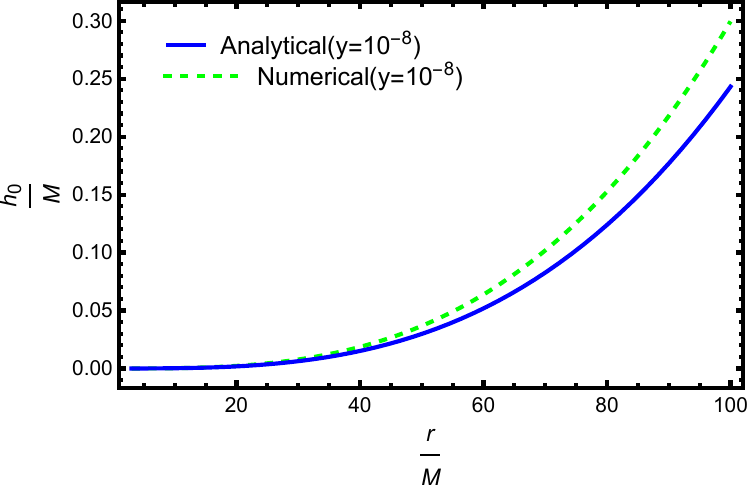}\\
\includegraphics[width=0.9\linewidth]{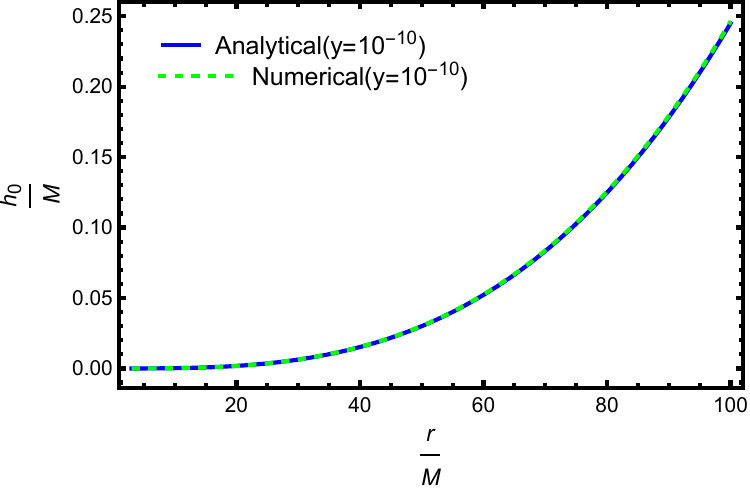}
\caption{In this figure we have compared the numerical and analytical solution for various values of $y$ keeping the values of $p$ and $k$ fixed. It is clear from the figure that as $y\rightarrow 0$, the numerical and analytical solution matches. In both the cases we have fixed $p=10^{-5}$ and $k=10^{-6}$.}\label{Numerical-investigation-down-modified-bc}
\end{figure}

Up to this point, we have only performed analytical computations. In this section, we will corroborate our analytical computation by directly solving the master equation numerically in Mathematica\cite{Mathematica}. For simplicity, here we have considered only Eq.[\ref{master-Love-2}]. To check our result, we have numerically integrated this equation with the previously discussed metric components of $R=0$ spacetime and imposing the boundary condition,  
\begin{align}
    h_{0}(r=2M)=0~,\label{bc-bh-1}\\
    h^{'}_{0}(r=2M)=k~\label{bc-2}.
\end{align}
Here $\prime$ denotes, derivative with respect to $r$ and $k$ is some finite constant (dimensionless). We call these boundary conditions as ``black hole-like boundary condition". This name can be justified by the fact that in the case of black holes the master variable $h_{0}$ automatically satisfies these boundary conditions (consider Eq.[\ref{h00}] for the case of black holes). Therefore, these boundary conditions are physically viable. Using these boundary conditions and taking $k=10^{-6}$ we have solved the master equation numerically up to some large radius $r=600M$. In Fig.\ref{Numerical-investigation-down} we have compared our analytical result with the numerical solution. In the \textit{top panel} of the figure we have compared the analytical and numerical solution of $h_{0}$ taking $p=10^{-3}$. And compared it with the corresponding black hole case. From the figure, it is clear that for large values of $r$ the analytical result and numerical results do not match. On the other hand, in the second and third panel of the figure we have done the same comparison for $p=10^{-5}$ and $p=10^{-10}$ respectively. It is clear that the analytical result Eq.[\ref{h0-down-analytical}] agrees with the numerical result for small values of $p$, that is for $p\gtrsim 0$. This is exactly the regime of working parameter space where the linear approximation Eq.[\ref{linear-master-2}] holds. Next we have numerically determined $k_{2}^{B}$(see Eq.[\ref{Magnetic-TLN2}]) by fitting the curve $A_{2}r^{3}+B_{2}r^{-2}$ with the numerical solution at large distance between $r=400M$ to $r=500M$ and determined $A_{2}$ and $B_{2}$ using the Mathematica's builtin package \texttt{BestFitParameters}. With working precession 40 in Mathematica, we get for the case of black holes, $k_{2}^{\rm BH}\sim O(10^{-10})$ which is vanishingly small. Then we repeat the same task for the wormhole case with $p=10^{-5}$ and find $k_{2}^{\rm WH}\sim O(10^{-10})$. This shows that both have same order order of magnitude within the working precession and therefore the conclusion of vanishingly small Love numbers of the wormholes is consistent with the analytical computations.
Similar consideration also holds for Eq.[\ref{master-Love}].\\
\indent We have further analyzed the sensitivity of these numerical solutions by slightly modifying the black hole-like boundary condition Eq.[\ref{bc-bh-1}]. That is, we wish to check what happens to analytical solutions with respect to the purely numerical solutions if we use the modified boundary condition:
\begin{align}\label{Modified-Bh-bc}
     &h_{0}(r=2M)=y M~,\\
      &h^{'}_{0}(r=2M)=k~
\end{align}
where $y$ is a dimensionless number and $y\gtrsim 0$. In Fig.\ref{Numerical-investigation-down-modified-bc} we have compared the numerical and analytical solutions for various values of $y$. It is clear from the figure that as $y\rightarrow 0$, the numerical and analytical solution (to Eq.[\ref{linear-master-2}]) matches. This indeed shows that the solution to the master equation is, in general, highly sensitive to any deviation from the corresponding black hole like case.
\section{Consideration of Damour-Solodukhin Wormhole}\label{sec-6}
To further check, if the results obtained above only works for the $R=0$ spacetime, we will analyse the case of Damour-Solodukhin wormhole\cite{Damour:2007ap,Bueno:2017hyj}. The line element is described by,
\begin{align}\label{DS-wormhole}
    ds^{2}=-\left(1-\frac{2M_{1}}{r}\right)dt^{2}+\frac{dr^{2}}{\left(1-\frac{2M_{2}}{r}\right)}+r^{2}d\Omega^{2}_{2}~.
\end{align}
Here $M_{1}=\frac{M_{2}}{(1+\delta^{2})}$ and $r=2M_{2}$ is the wormhole throat. 
Like the previous case of the $R=0$ spacetime, here also, we will consider $\delta^{2}\gtrsim 0$\footnote{Here $\delta^{2}$ is a regularization parameter.}, so the phenomenologically the wormhole spacetime mimics the black hole spacetime. It can be shown that the spacetime is sourced by anisotropic fluid whose energy momentum tensor has the following components,
\begin{align}\label{DS-EM}
    &\rho=0,~p_{r}=-\frac{\delta ^2 M_{2}}{4 \pi  r^2 \left(-2 M_{2}+\delta ^2 r+r\right)}+O(\delta^{4})\nonumber\\
   & p_{t}=\frac{\delta^{2}  M_{2} (r-M_{2})}{8 \pi  r^2 (r-2 M_{2})^2}+O(\delta^{4})~.
\end{align}
To check the validity of our algorithm, as discussed in Sec.\ref{sec-4}, we will consider Eq.[\ref{master-Love-2}]. It turns out that, up to ${O}(\delta^{2})$\footnote{Here $\delta^{2}$ should be treated as an expansion parameter and we have kept terms linear in $\delta^{2}$.}, we have,
\begin{align}\label{DS-master-down}
   \mathcal{O}[h_{0}^{(0)}]+\delta^{2}\mathcal{O}[h_{0}^{(1)}]=\delta^{2}S_{D}[r]~.
\end{align}
This time the operator $\mathcal{O}$ is obtained by replacing $M$ by $M_{2}$ in Eq.[\ref{Operator-O}]. For brevity, we have omitted the expression of the source term $S_{D}$. One can use the solution Eq.[\ref{h00}] to obtain the value of the source term $S_{D}[r]$ as,
\begin{align}
    S_{D}[r]=AM_{2}.
\end{align}
With these one obtains the following solution for $h_{0}^{(1)}$ (exactly!):
\begin{align}
    h_{0}^{(1)}[r]=&-\frac{A r^3}{8}+q_{1} r^2 (r-2 M_{2})\\
    &-\frac{q_{2}r^{4}}{8 M^{5}_{2}r}\bigg(1-\frac{2M_{2}}{r}\bigg)\log\bigg(1-\frac{2M_{2}}{r}\bigg)\nonumber\\
    &+\frac{q_{2}(4M_{2}^{4}+4M_{2}^{3}r+6M_{2}^{2}r^{2}-6M_{2}r^{3})}{24 M_{2}^{5}r}~.
\end{align}
Here $q_{1}$ and $q_{2}$ are arbitrary constants. Now regularity of $h_{0}^{(1)}[r]$ at the wormhole throat demands, $q_{2}=0$. And in this context the total solution becomes,
\begin{align}\label{h0-down-total-ds}
    h_{0}[r]=-\frac{A \delta^{2} r^3}{8}+(A+\delta^{2} q_{1}) r^{3} \bigg(1-\frac{2 M_{2}}{r}\bigg)~.
\end{align}
Like the claim we made in Sec.\ref{sec-4}, here also we do not have any $\frac{1}{r^{2}}$ term. Then it follows that the axial tidal Love number vanishes perturbatively. Moreover, in this case also $h_{0}(r=2M_{2})=0$ implies $A=0$ and vice versa.

\section{Conclusion}\label{sec-7}
Tidal Love numbers characterise the induced multipole moments\cite{RevModPhys.52.299}, that is, the response of a compact object under the influence of an external tidal field\cite{Cardoso:2025npr}. Depending on the parity of the external perturbation, the tidal Love numbers can be divided into two parts\cite{Binnington:2009bb}: a) magnetic type and b) electric type. Tidal deformations of an object can be detected by ground based detectors during the inspiral phase of the binary merger. This can give us a hint about the nature of the compact object or the underlying theory of gravity as it is shown that the tidal Love number of black holes in general relativity vanishes identically. Moreover, it turned out that in the case of a neutron star, the electric type tidal Love number decreases monotonically as the compactness increases and vanishes in the black hole limit. On the other hand, the magnetic type tidal Love number vanishes when the compactness goes to zero, showing that it has no Newtonian analogue. \\
\indent In this paper, we have studied the tidal Love number of wormhole spacetimes\cite{Battista:2024gud,DeFalco:2021ksd,DeFalco:2021klh,DeFalco:2020afv,DiGrezia:2017daq}. Wormholes are fascinating alternatives to black hole geometries, which connects two distinct universes using a bridge like structure, called wormhole throat. In general, wormhole spacetime metric comes with additional parameters as compared to their black hole counterpart which are described by only three hairs, namely mass, spin and charge. The job of the additional parameter is to regularise\cite{Simpson:2018tsi,Damour:2007ap,Franzin:2022iai} the geometry in the sense that there should not be any curvature singularity. Here, we have analysed the perturbation of a special wormhole solution in the context of $R=0$ spacetime. This wormhole geometry contains $p$ as regularisation parameter. We have discussed axial gravitational perturbations in the context of GR (see also \cite{Biswas:2022wah}). In particular, we have focused on the strictly static ($\omega=0$) perturbations. Eq.[\ref{master-Love}] and Eq.[\ref{master-Love-2}] are the associated master equations, and they depend on whether one chooses $u^{\mu}$ or $u_{\mu}$ respectively as a fundamental quantity while considering the fluid perturbations. Interestingly enough, it turns out that these two equations take the same form in the case of vacuum spacetime. We solve both of these equations perturbatively in the regularisation parameter $p$ and keep terms up to linear order. Further, we have imposed the condition on the solution that it must be regular at the wormhole throat. With these, we have asked for the behaviour of the solution at large $r$. It turns out that the coefficient of $\frac{1}{r^{2}}$ term vanishes, which leads us to conclude that under strictly static perturbation of this wormhole geometry, the magnetic-type tidal Love number vanishes (perturbatively up to linear order in regularisation parameter $p$). The careful reader must distinguish this perturbatively vanishing Love number from that of the vanishing Love number of black holes in general relativity, as the latter is an exact result. Our result also justifies the title of the paper in the sense that the wormhole spacetime under consideration is a very good mimicker of the black hole spacetime under perturbative study.~Further in Sec.\ref{sec-5}, we have corroborated our result by numerically by solving the master equation. In Sec.\ref{sec-6} we have also extended this calculations towards the case of Damour-Solodukhin wormhole and obtained similar behavior.\\ 
\indent As a future direction, we wish to study the polar perturbation of this geometry elsewhere. Moreover, it is shown\cite{Hui:2021vcv,Hui:2022vbh,Hioki:2008zw,BenAchour:2022uqo,Charalambous:2021kcz} that the existence of ladder symmetry in the master equation gives rise to the vanishing of the Love number of  black holes. The status of this ladder symmetry argument is still unrevealed in the context of wormhole geometries. Such a study in the context of wormholes will be interesting. Further, it is shown in\cite{DeLuca:2024ufn} that the tidal response of black holes can be understood in terms of the Green's function associated with wave propagation. Since it is known\cite{Mark:2017dnq} that the Green's function of ECOs can be written in terms of the black hole's Green's function and using the transfer function, it would be interesting to check if the tidal response of the wormhole is indeed captured by the transfer function. On the other hand, the tidal Love numbers of these wormhole geometries under fermionic and electromagnetic perturbations can be considered as an interesting extension of this work. 

\section*{Acknowledgement}
Research of SB is supported by the NPDF fellowship provided by ANRF file no:~PDF/2025/002040. SB thanks Sumanta Chakraborty for various helpful discussions. SB would like to thank Sayan Chakrabarti for reading the manuscript. SB also thanks IACS, IITGN, IUCAA and Sapienza University of Rome for providing hospitality during which part of this work was conducted. SB would like to thank ICTS for providing hospitality during the conference ``Future of Gravitational-Wave Astronomy" (code: ICTS/fgwa2025/10). SB would like to thank ICTS for providing hospitality, as a part of this work is done during ``Summer School on Gravitational Wave Astronomy-2026". 
\appendix
\section{S[r] approximated near the throat}\label{app-A}
Here we will demonstrate the solution of the master equation Eq.[\ref{master-Love}] when the inhomogeneous term $S[r]$ is approximated near the throat at $r=2M$. Since the inhomogeneous term Eq.[\ref{linear-source}] contains a surd term in the denominator, so it's behavior near the throat will contain non-polynomial behavior. To avoid such problem we use the following approximation for this term. For that purpose we note that 
\begin{align}\label{Key-approximation-1}
  \sqrt{1-\frac{2M}{r}}=\sqrt{\left(1-\frac{M}{r}\right)^{2}-\frac{M^{2}}{r^{2}}}~.
\end{align}
Now for the near throat approximation $r\gtrsim 2M$ we can ignore the last term of the above expression, to yield
\begin{align}\label{Key-approximation-2}
  \sqrt{1-\frac{2M}{r}}\approx \left(1-\frac{M}{r}\right) 
\end{align}
With this the inhomogeneous term can be further approximated as,
\begin{align}\label{source-near-thraor-1}
  S[r]\approx A M \left(1-\frac{4M}{r}\right)\left(1-\frac{M}{r}+\frac{M^{2}}{r^{2}}\right). 
\end{align}
Then the solution to $\mathcal{O}h_{0}^{(1)}=S[r]$ can be computed without handling any mathematical complexity, which is
\begin{align}\label{near-sol-h1}
    &h_{0}^{(1)}(r)=c_{1}r^{3}\left(1-\frac{2M}{r}\right)+\left(\frac{39 A} {72}-\frac{3 c_{2}}{24 M^{5}}\right)r^{3}\sigma\nonumber\\
    &+\frac{c_{2}}{24 r M^{5}}(4 M^{4}+4 M^{3}r+6M^{2}r^{2}-6Mr^{3})\nonumber\\
    &+\frac{A}{72}(58r^{3}-56M r^{2}-18M^{2}r-72M^{3})~.
\end{align}
Here we have defined $\sigma(r)\equiv\left(1-\frac{2M}{r}\right)\ln{\left(1-\frac{2M}{r}\right)}$. Like the previous cases one can realize that the arbitrary constants $c_{1}$ and $c_{2}$ are coming due to the solution of the homogeneous equation $\mathcal{O}h_{0}^{(1)}=0$ and the terms involving $A$ are coming due to the particular integral. In this case also the regularity of the solution at the throat gives,
\begin{align}\label{regularity-near}
39 A M^{5} = 9 c_{2}~.    
\end{align}
Next, we can analytically continue the above solution to the large values of $r$ and show that due the condition Eq.[\ref{regularity-near}] we cannot have $\frac{1}{r^{2}}$ term in the expansion. In other words, the tidal Love number vanishes.  One may argue that the condition Eq.[\ref{regularity-near}] appears due to the approximation Eq.[\ref{Key-approximation-2}] and try to compute the the solution without the approximation. It can be shown that such a solution contains non-polynomial behavior or $r$ when expanded to large $r$. Interestingly, no $\frac{1}{r^{2}}$ term will be present in the expansion and the Love number vanishes. Due it's complicated nature we avoid it writing here. This is how a simple approximation Eq.[\ref{Key-approximation-2}] makes life easier while keeping the physical part intact. 

\section{Adding free scalar field}\label{app-B}
Here we discuss the scalar perturbation of the $R=0$ wormhole spacetime. We add a free massless test scalar field to the geometry, which is described by,
\begin{align}\label{scalar field}
    g^{\mu\nu}\nabla_{\mu}\nabla_{\nu}\Phi=0~.
\end{align}
Since the background spacetime is static and spherically symmetric, we can decompose the scalar field as follows,
\begin{align}\label{Scalar-decomposition}
\Phi(t,r,\theta,\phi)=\sum^{\infty}_{l=0}\sum_{m=-l}^{l}\frac{\Psi_{\ell m}(r)}{r}Y_{\ell m}(\theta,\phi)e^{-i\omega t}~.
\end{align}
Then putting this ansatz in Eq.[\ref{scalar field}] and using the properties of $Y_{\ell m}(\theta,\phi)$ we can get the following master equation\cite{Dey:2020lhq}
\begin{align}\label{scalar-maste equation}
    \sqrt{-g_{tt}g^{rr}}\partial_{r}&\left(r^{2}\sqrt{-g_{tt}g^{rr}}\partial_{r}\chi_{\ell m}\right)\nonumber\\
    &+\left(r^{2}\omega^{2}-\ell(\ell+1)\right)\chi_{\ell m}=0~,
\end{align}
where $\chi_{\ell m}=\frac{\Psi_{\ell m} }{r}$~. Here also, we solve the equation in the static case assuming a linear expansion of the master variable $\chi_{\ell m}$ in $p$, namely $\chi=\chi_{(0)}(r)+p \chi_{(1)}(r)$. It turns out that 
\begin{align}\label{Linear-scalar}
   \mathcal{O}_{s}\chi_{(0)}+p\mathcal{O}_{s}\chi_{(1)}=S_{s}[r]~. 
\end{align}
Here the operator $\mathcal{O}_{s}$ has the form,
\begin{align}\label{scalar-operator}
  \mathcal{O}_{s}\equiv  r (r-2 M) \frac{d^{2}}{dr^{2}}+2 (r-M) \frac{d}{dr}-\ell (\ell+1),
\end{align}
and the source term $S_{s}[r]$ is simply given by,
\begin{align}\label{scalar-source}
    S_{s}[r]\equiv \frac{M}{\sqrt{1-\frac{2M}{r}}}\frac{d\chi_{(0)}}{dr}~. 
\end{align}
Insisting on the interesting phenomenological case of $\ell=2$ and regularity on the throat $r=2M$, we can get the following solution,
\begin{align}\label{l=2-leading order-scalar}
    \chi_{(0)}(r)= A\left(\frac{2M^{2}-6Mr +3 r^{2}}{2 M^{2}}\right)~,
\end{align}
where $A$ is an arbitrary constant. Using this leading order solution we obtain the following solution for $\chi_{(1)}(r)$:
\begin{align}\label{l=2-NLO-scalar}
   & \chi_{(1)}(r)=C_{1}\left(\frac{2M^{2}-6Mr +3 r^{2}}{2 M^{2}}\right)+\nonumber\\
   &C_{2}\left[\frac{3}{2}\left(1-\frac{r}{M}\right)+\left(\frac{2M^{2}-6Mr +3 r^{2}}{4 M^{2}}\right)\log\left(\frac{r}{r-2M}\right)\right]\nonumber\\
   &+A (...)~.
\end{align}
Here $C_{1}$ and $C_{2}$ are arbitrary constants.
For brevity, we have omitted the expression for the terms involving the last expression $A(...)$. One can check using Mathematica, that it involves terms having $\left(\log(1-\frac{2M}{r})\right)^{2}$, that is, to get regular solution at the throat we must impose $A=0=C_{2}$. This is similar to the case of an axial-gravitational perturbation. Therefore, the total solution is,
\begin{align}\label{}
    \chi(r)=p C_{1}\left(\frac{2M^{2}-6Mr +3 r^{2}}{2 M^{2}}\right)~.
\end{align}
Contrary to the case of axial gravitational perturbation, the scalar perturbation is of even parity type, so the leading order growing part at larger $r$ is $r^{\ell }$ and the dominant decaying part is $r^{-(\ell+1)}$. Therefore, from the above expression it is clear that in this case we have only growing part at large $r$. That is under scalar perturbation the Love number vanishes (perturbatively).
\bibliography{Axial-TLN}
\bibliographystyle{utphys1}

\end{document}